\documentclass[aps,prx,groupedaddress,twocolumn,10pt, superscriptaddress, reprint]{revtex4-2}

\usepackage{lipsum}

\usepackage{amsmath}

\usepackage{amssymb}

\usepackage{amsthm}

\usepackage{bbm}

\usepackage{graphicx}

\usepackage{color}

\usepackage{upgreek}

\usepackage[linkcolor = blue, citecolor = blue, urlcolor = blue, colorlinks = true]{hyperref}

\usepackage{amssymb}

\usepackage{bm}

\usepackage[capitalise]{cleveref}

\usepackage{textcomp}

\usepackage{hyperref}

\usepackage[usenames,dvipsnames]{xcolor}

\usepackage[normalem]{ulem}

\usepackage{wasysym}

\usepackage{tikz}
\hypersetup{colorlinks=true, linkcolor=BrickRed, urlcolor=blue!50!black, citecolor=blue!50!black}

\newcommand\diff{\mathrm{d}}

\renewcommand{\vec}[1]{\mathbf{#1}}
\renewcommand{\phi}[0]{\varphi}

\usepackage[normalem]{ulem}

\begin{document}

\title{Crowding-Enhanced Diffusion: An Exact Theory for Highly Entangled Self-Propelled Stiff Filaments}
\date{\today}
\def\hhu{\affiliation{Institut f\"ur Theoretische Physik II: Weiche Materie, Heinrich-Heine Universit\"at D\"usseldorf, Universit\"atsstr.~1, 40225 D\"usseldorf, Germany}}
\def\inn{\affiliation{Institut f\"ur Theoretische Physik,
  Universit\"at Innsbruck, A-6020 Innsbruck, Austria}}
\def\tud{\affiliation{Theorie Weicher Materie, Fachbereich Physik, Technische Universität Darmstadt, Hochschulstraße 12, 64289 Darmstadt, Germany}}

\author{Suvendu Mandal}
\thanks{S.M. and C.K. contributed equally.}
\email{\newline suvendu.mandal@pkm.tu-darmstadt.de}
\affiliation{Institut f\"{u}r Theoretische Physik II: Weiche Materie, Heinrich-Heine-Universit\"{a}t D\"{u}sseldorf, D-40225 D\"{u}sseldorf, Germany}
\author{Christina Kurzthaler}
\thanks{S.M. and C.K. contributed equally.}
\email{\newline ckurzthaler@pks.mpg.de}
\affiliation{Department of Mechanical and Aerospace Engineering, Princeton University, Princeton, NJ 08544, USA}
\affiliation{Institut f\"ur Theoretische Physik,
	Universit\"at Innsbruck, Austria}
\author{Thomas Franosch}
\email{thomas.franosch@uibk.ac.at}
\affiliation{Institut f\"ur Theoretische Physik,
	Universit\"at Innsbruck, Austria}
\author{Hartmut L\"owen}
\email{hlowen@hhu.de}  
\affiliation{Institut f\"{u}r Theoretische Physik II: Weiche Materie, Heinrich-Heine-Universit\"{a}t D\"{u}sseldorf, D-40225 D\"{u}sseldorf, Germany}

\begin{abstract}
We study a strongly interacting crowded system of self-propelled stiff filaments by event-driven Brownian dynamics simulations and an analytical theory to elucidate the intricate interplay of crowding and self-propulsion. We find a remarkable increase of the effective diffusivity upon increasing the filament number density by more than one order of magnitude. This counter-intuitive \emph{'crowded is faster'} behavior can be rationalized by extending the concept of a confining tube pioneered by Doi and Edwards for highly entangled crowded, passive to active systems. We predict a scaling theory for the effective diffusivity as a function of the P\'eclet number and the filament number density. Subsequently, we show that an exact expression derived for a single self-propelled filament with motility parameters as input can predict the non-trivial spatiotemporal dynamics over the entire range of length and time scales. In particular, our theory captures short-time diffusion, directed swimming motion at intermediate times, and the transition to complete orientational relaxation at long times.
\end{abstract}

\maketitle

The cytoskeleton composed of various biofilaments is a prime example of a strongly interacting, crowded system and represents a prerequisite building block of all living cells. A distinguishing feature of these biofilaments is their ability to  self-propel in the presence of motor proteins~\cite{Howard:2001mechanics,Albert:2008molecular,Schaller:2010,Bausch:2006,Sanchez:2012}.
Their individual transport properties inside the cytoskeleton play a crucial role for the proper functioning of the cell including its migration and mitosis and, thus, provide an inevitable ingredient for the design of biology-inspired materials, e.g., synthetic cells~\cite{Blain:2014,Ganzinger:2019}. Yet, a theory for these highly entangled out-of equilibrium systems remains a challenge and poses a complex problem already at the single-particle level, where the swimming direction of the active agent is strongly dictated by obstacles~\cite{Chepizhko:2013,Contino:2015,Reichhardt:2017negative},
attractive traps~\cite{Pototsky:2012active,Takatori:2016acoustic}
and topological constraints~\cite{Volpe:2017topography,Souslov:2017topological}. Thus, the development of  analytical theories is important for our future understanding of transport processes in biological systems, such as the interior of cells, soils, and biofilms, and the design of novel nanotechnological devices, including targeted drug delivery~\cite{Li:2017, Sedighi:2019} and bioremediation tools~\cite{Soler:2013,Gao:2014,Adadevoh:2016,Ren:2019}.

Experiments of artificial Janus particles~\cite{Morin:2017PRE,Brown:2016swimming,Takagi:2014hydrodynamic,Morin:2017NaturePhysics},
bacteria~\cite{Brown:2016swimming,Bhattacharjee:2019bacterial,Frangipane:2019invariance,Makarchuk:2019enhanced}
and active granulates~\cite{Kumar:2014flocking} in a fixed obstacle matrix have revealed a drastic change of the dynamics compared to their motion in a free environment. These include localization by strong disorder~\cite{Kumar:2014flocking,Morin:2017PRE}, hydrodynamic trapping~\cite{Takagi:2014hydrodynamic} or scattering~\cite{Contino:2015}, and trajectories reminiscent of L{\'e}vy walks~\cite{Bhattacharjee:2019bacterial,Frangipane:2019invariance}. Computer simulations of basic model systems in complex environments have been performed under various conditions~\cite{Chepizhko:2013,Chepizhko:2013Optimal,Quint:2015topologically,Reichhardt:2014disorderlandscapes,Zeitz:2017active,Chepizhko:2019ideal} and reveal remarkable phenomena ranging from clogging and depinning~\cite{Reichhardt:2018clogging},
negative differential mobility~\cite{Reichhardt:2017negative}, to
new dynamical scaling laws~\cite{Mokhtari:2019,Morin:2017NaturePhysics}. 

Analytical models on active transport in crowded environments~\cite{Bertrand:PRL2018,Illien:2016, Alonso-Matilla:2019}, however, rely on periodic or lattice structures and neglect the anisotropic feature of many self-propelled agents, such as biofilaments and bacteria.
The diffusive transport of stiff biofilaments in a disordered matrix has been investigated  using an analytical theory and computer simulations~\cite{Hoefling:2008,Leitmann:2016}, but the effects due to activity have not been explored yet. 
Thus, analytical theories which incorporate
both, self-propulsion of anisotropic agents and the nature of the disordered environment, 
could serve as a paradigmatic model system of statistical mechanics.

In this Letter, we use event-driven Brownian dynamics simulations and an analytical theory to study a strongly interacting system of a dense solution of self-propelled stiff filaments. First, our computer simulations reveal that crowding can enhance their long-time effective diffusion by more than one order of magnitude. We explain this counterintuitive behavior by extending the concept of a confining tube pioneered by Doi and Edwards~\cite{Doi:1978tube} to active solutions and  present a scaling theory for the competition between self-propulsion and crowding. Second, and strikingly, we provide an exact analytical expression of the intermediate scattering function for a single self-propelled filament in a densely, crowded dynamic environment with measured motility parameters as input, in the limit of high entanglement.
Therefore, the active needle suspension constitutes a rare example of a strongly interacting non-equilibrium system allowing for a complete analytic solution.

\paragraph*{Model.--}
We investigate semidilute solutions of $N$ self-propelled stiff filaments within a cubic box of volume $V$  in $3$D, each filament having length $L$ and diameter $b$ much smaller than its length, $b/L\to0$, so that it is approximated by an infinitely thin needle. Crowded  solutions of self-propelled filaments remain globally isotropic even
at higher densities, $n=N/V$, as the transition from isotropic to nematic order occurs at $nL^2b\sim \mathcal{O}(1)$ corresponding to $n\to\infty$ for infinitely thin needles~\cite{Onsager1949:effects}. Hydrodynamic interactions can be neglected, as infinitely thin needles cannot drag fluid. Therefore, the dynamics are dictated by the topological constraints that the needles cannot cross each other, which we account for in a  pseudo-Brownian scheme~\cite{Leitmann:2016,Leitmann:2017PRE}. We determine possible collisions during every Brownian time step, $\tau_B$, and update their positions and orientations by enforcing conservation of energy and (angular) momentum (see SI~\cite{suppl}). 
Between two collisions, the active agent moves at a constant velocity $v$ along its instantaneous orientation $\vec{u}$, which performs rotational diffusion characterized by the diffusivity $D_\text{rot}
^0$. It is subject to short-time anisotropic translational diffusion with coefficients parallel ($D_{\parallel}^{0}$) and perpendicular ($D_{\perp}^{0}$) to the filament. The equations of motion for the position, $\vec{r}$, and the orientation, $\vec{u}$, of the active agent (between the collisions) read
\begin{align}\label{eq:langevin}
\frac{\diff \vec{r}}{\diff t} &=v\vec{u} + \Big[ \sqrt{2 D_\parallel^0} \vec{uu} + \sqrt{2 D_\perp^0} (\mathbb{I} - \vec{uu}) \Big]\boldsymbol\eta, \\
\frac{\diff \vec{u}}{\diff t} &= -2D_\text{rot}^0 \vec{u} - \sqrt{2D_\text{rot}^0} \vec{u} \times \boldsymbol\xi, \label{eq:u}
\end{align}
where 
$\boldsymbol{\eta}$ and $\boldsymbol{\xi}$ represent independent Gaussian white noise with zero mean of unit strength. The rotational diffusivity sets the bare rotational relaxation time $\tau_\text{rot}^{0}=1/2D_\text{rot}^0$~\cite{Berne2000dynamic}. The short-time transport coefficients are not independent, rather hydrodynamics for long rods entails $D_\parallel^0=2D_\perp^{0}$ and $D_\text{rot}^{0} = 12 D_\perp^0/L^2$~\cite{DoiEdwards:1988}. Therefore, the behavior of the system is controlled by two dimensionless numbers: the (reduced) number density $n^\star:=nL^{3}$ and the P\'{e}clet number $\text{Pe}:=v L/\bar{D}^0$ measuring the strength of self-propulsion with respect to diffusion with average diffusivity $\bar{D}^0=\left(D_\parallel^0+2D_\perp^0\right)/3$. 

\begin{figure}[htp]
\includegraphics[width=\linewidth]{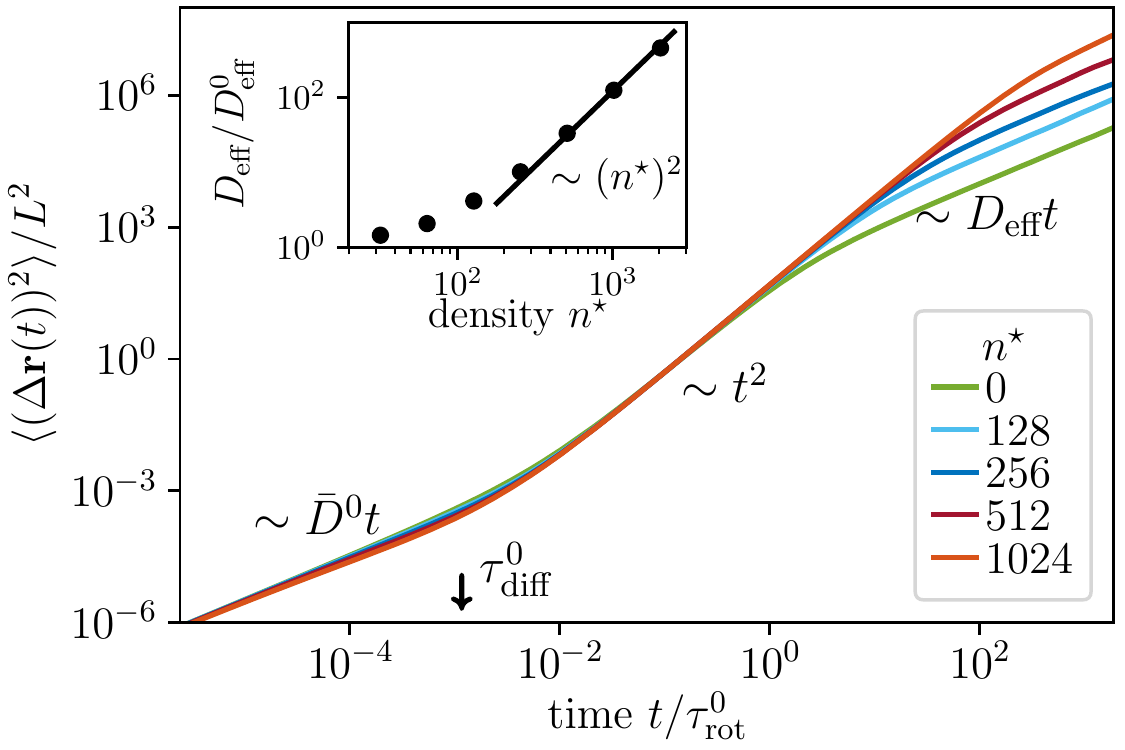}
\caption{\label{fig:enhanced_diffusion}
Mean-square displacement, $\langle(\Delta\vec{r}(t))^2\rangle$, of a  self-propelled stiff filament suspended in needle solutions with different number densities $n^\star=nL^3$. Here, 
the P{\'e}clet number is $\text{Pe}=125$. (\emph{Inset}) Effective diffusivity, $D_\text{eff}$, as function of the number density $n^\star$ rescaled by the effective diffusivity in a free environment, $D_\text{eff}^0$. The solid black line indicates a quadratic increase.
}
\end{figure}
\paragraph*{Enhanced effective diffusion.--}
To quantify the effect of crowding, we investigate the mean-square displacement as a function of the number density, $n^\star$, at a fixed P\'eclet number $\text{Pe}=125$ [Fig.~\ref{fig:enhanced_diffusion}]. At short times, $t\lesssim\tau_\text{diff}^0:=\bar{D}^0/v^2$, crowding plays no role and filaments are  diffusive ($\sim \bar{D}^0t$), while at intermediate times directed motion ($\sim t^2$) dominates, and eventually reaches the terminal diffusive regime at long times. Most prominently, as the number density increases, the long-time diffusion increases drastically. This means that, counter-intuitively, the diffusivity of an individual self-propelled filament can  be enhanced by adding more self-propelled filaments to the system. To quantify this phenomenon, we determine the long-time effective 
diffusivity $D_{\text{eff}} := \lim_{t\to \infty} \langle (\Delta \vec{r}(t))^2\rangle/6 t$, and present normalized $D_\text{eff}/D_\text{eff}^0$ as a function of $n^\star$ in the inset of Fig.~\ref{fig:enhanced_diffusion}, where $D_\text{eff}^0=\bar{D}^0+v^2\tau_\text{rot}^0/3$ is the effective diffusivity at infinite dilution~\cite{Kurzthaler:2016ScRprt}. For $n^\star \gtrsim 100$, an increase of the effective diffusion is observed by more than one order of magnitude. Most significantly, the dependence of  $D_\text{eff}/D_\text{eff}^0$ as a function $n^\star$ suggests a power-law scaling $D_\text{eff}/D_\text{eff}^0 \sim (n^\star)^2$. We rationalize this enhanced effective diffusion qualitatively via the following argument. As the solution becomes more and more crowded, these self-propelled filaments are forced to move within an effective tube formed by their neighboring filaments, leading to an increased rotational relaxation time $\tau_\text{rot}$. Consequently, the swimming directions are preserved for  long times, which is evident by the extended directed regime ($\sim t^2$) with increasing density [Fig.~\ref{fig:enhanced_diffusion}]. Further we anticipate that crowding-enhanced diffusion originates from an increased $\tau_\text{rot}$, i.e., $D_\text{eff}/D_\text{eff}^0 \sim \tau_\text{rot}/\tau_\text{rot}^0$, at fixed P\'eclet number \text{Pe}.

\begin{figure}[htp]
\includegraphics[width=\linewidth]{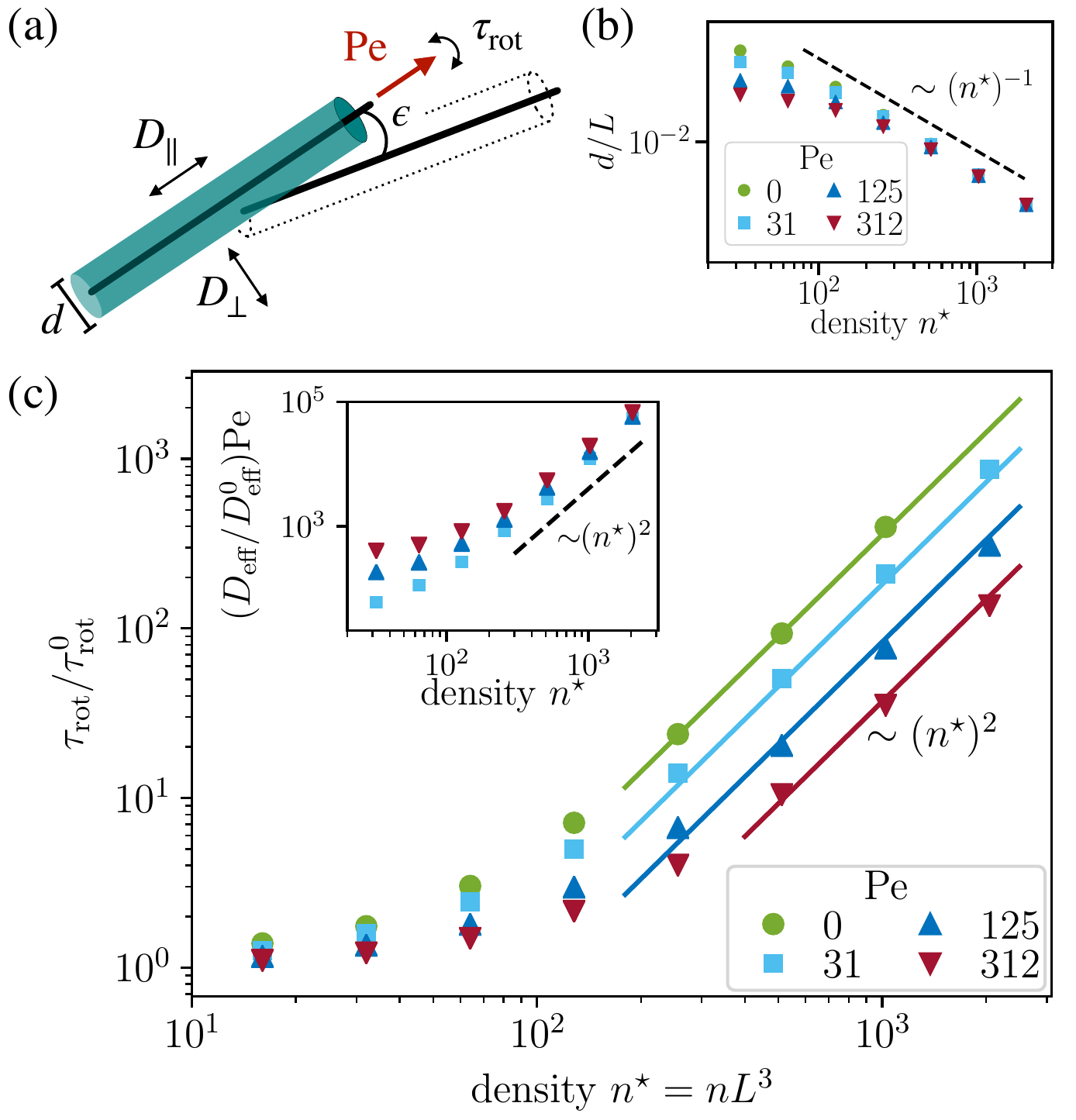}
\caption{\label{fig:cartoon_plus_rotation}
(a) In the highly entangled regime (\emph{mesoscale}) a self-propelled filament is confined to an effective tube of diameter $d$ created by neighboring self-propelled filaments with motility parameters $\text{Pe}$, $D_\perp$, $D_\parallel$, and $\tau_\text{rot}$. At a later time, the filament is confined to a new tube tilted by an angle $\epsilon$ with respect to the initial tube. (b) Tube diameter $d$ as a function of density $n^\star$. (c) Orientational relaxation time $\tau_\text{rot}$ extracted from simulations at different number densities $n^\star$ for several P{\'e}clet numbers. (\emph{Inset}) Rescaled effective diffusivities, $D_\text{eff}$, for finite P{\'e}clet numbers.
}
\end{figure}

\begin{figure*}[tp]
    \includegraphics[width=\textwidth]{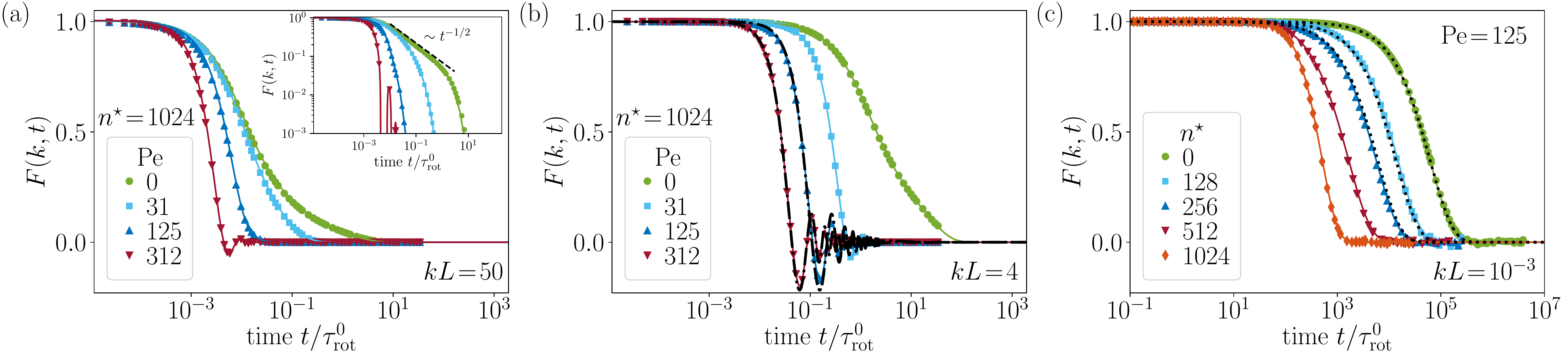}
	\caption{\small Intermediate scattering function, $F(k,t)$, for (a-b) a broad range of P\'{e}clet numbers $\text{Pe}$ at fixed density $n^\star=1024$ and (c) different densities $n^\star$ at fixed P\'{e}clet number $\text{Pe}=125$. The solid lines are the theoretical predictions of the tube model [Eq.~\eqref{eq:isfanalytic}], and symbols are from simulations. Panels (a),(b), and (c) represent data for different wave numbers $kL=50, 4$, and $10
^{-3}$, respectively, where $L$ denotes the filament length. The inset in (a) shows the data in log-log scale, where the dashed solid line indicates an algebraic decay $\sim t^{-1/2}$.  The black dotted-dashed lines in (b) indicate the sinc-function, $\sin(vkt)/(vkt)$, and the dotted lines in (c) represent effective diffusion, $\exp(-k^2D_\text{eff}t)$.}
	\label{fig:ISF}
\end{figure*}

\paragraph*{Validity of the tube theory.--} To obtain more insight into the scaling behavior of $D_\text{eff}/D_\text{eff}^0$, we measure the topological constraints felt by the self-propelled filament. In particular, we investigate the time-dependent orientational correlation function $\langle \vec{u}(t)\cdot \vec{u}(0)\rangle$ for a range of P\'{e}clet numbers, $\text{Pe}=0,\dots 312$ (see SI~\cite{suppl}). It turns out that the shape of the relaxation becomes purely exponential as a function of time, $\exp(-t/\tau_\text{rot})$, characterized by the long-time rotational relaxation time $\tau_\text{rot}$. The rotational relaxation times extracted from the simulations depend on the P{\'e}clet number and the number density [Fig.~\ref{fig:cartoon_plus_rotation}(c)]. In particular, they follow an asymptotic power law, $\tau_\text{rot} \sim (n^{\star})^{2}$, for all P\'{e}clet numbers considered. This provides a surprisingly robust explanation for the enhanced effective diffusion due to crowding, i.e., $D_\text{eff}/D_\text{eff}^0 \sim \tau_\text{rot}/\tau_\text{rot}^0 \sim (n^\star)^2$. In fact, Fig.~\ref{fig:cartoon_plus_rotation}(c)suggests that the highest P\'{e}clet number $\text{Pe}=312$ studied here can decrease the prefactor of the asymptotic scaling by a factor of $\sim 11$ compared to the passive case ($\text{Pe}=0$).

We rationalize the asymptotic power law for the active crowded solution, $\tau_\text{rot} \sim (n^\star)^2$, with a tube concept for an active needle system. The tube model has been introduced by Doi-Edwards for passive crowded systems~\cite{Doi:1978tube} and describes the dynamics of entangled solutions, where every infinitely thin rod is confined to a tube of diameter $d$ created by its neighboring rods [Fig.~\ref{fig:cartoon_plus_rotation}(a)].
Following Doi-Edwards~\cite{Doi:1978tube}, the tube diameter which measures the transverse motion of the filament can be estimated as $d \sim 1/nL^2$, see SI~\cite{suppl}. This predicted scaling, $d/L \sim (n^\star)^{-1}$, remains valid for all P\'eclet numbers, yet upon close inspection the range of validity drifts to larger number densities with increasing P\'eclet number [Fig.~\ref{fig:cartoon_plus_rotation}(b) and SI~\cite{suppl}]. 
We further introduce the disengagement time as the time the agent requires to move its own length, $L$, along the tube. Since the transverse motion is essentially frozen,  the filament self-propels and diffuses freely along the tube at time scales $L/v$ and $L^2/D_\parallel^0$, respectively. The disengagement time is controlled by the faster of these two mechanisms and we use as estimate the interpolation formula  $\tau_0^{-1}= D_\parallel^0/L^2+v/L$.
In addition to motion along the tube-direction, the filament slightly rotates while entering a new tube by an angle $\epsilon\sim d/L$ with respect to the initial tube [Fig.~\ref{fig:cartoon_plus_rotation}(a)]. Thus, we find the asymptotic power-law for the rotational relaxation time
\begin{equation}
	\tau_\text{rot} \sim \frac{1}{\epsilon^2\tau_0^{-1}} \sim \frac{(n^\star)^2}{( D_\parallel^0/\bar{D}^0 + \text{Pe})} \tau_\text{rot}^0 ,
\end{equation}
which recovers the scaling law of Doi-Edwards for passive systems, $\tau_\text{rot}/\tau_\text{rot}^0\sim (n^\star)^2$, and reduces to $\tau_\text{rot}/\tau_\text{rot}^0\sim(n^\star)^2\text{Pe}^{-1}$ for large P{\'e}clet numbers.
Accordingly, the effective diffusivity can be predicted as
\begin{align}
    \frac{D_\text{eff}}{D_\text{eff}^0}\sim \frac{\tau_\text{rot}}{\tau_\text{rot}^0} \sim \frac{(n^\star)^2}{\text{Pe}},
\end{align}
which is confirmed by an asymptotic data collapse in the regime of high entanglement and P\'eclet number $\text{Pe} \gtrsim 50 $  [Fig.~\ref{fig:cartoon_plus_rotation}(c)~(\emph{inset})]. Our analysis demonstrates that the rotational relaxation time, $\tau_\text{rot}$, and the long-time effective diffusion, $D_\text{eff}$, of a self-propelled filament in a crowded environment indeed obey the scaling predictions of the tube theory, which provides insights into the underlying microscopic dynamics.

\paragraph*{Exact spatiotemporal dynamics.--}
The validity of the tube concept suggests a physical situation sketched in Fig.~\ref{fig:cartoon_plus_rotation}(a), which allows us to map a strongly interacting many-body problem onto an effective tube theory at the mesoscale. The key idea is to solve for the dynamics of a single self-propelled stiff filament in free space and use the motility parameters $\text{Pe}$, $D_\parallel= D_\parallel (n^\star)$, $D_\perp=D_\perp(n^{\star})$, $\tau_\text{rot}=\tau_\text{rot}(n^{\star})$ measured from the simulations as input to predict the full spatiotemporal dynamics of a filament immersed in a highly entangled system. 

To explore the full ramifications of the effective tube model, we compute the intermediate scattering function (ISF), characterizing the motion in space and time, 
\begin{equation}
\label{eq:genisf}
F(k,t)= \langle \exp[-i \vec{k}\cdot \Delta\vec{r}(t)] \rangle,
\end{equation}
where $\Delta \vec{r}(t)=\vec{r}(t) -\vec{r}(0)$ is the displacement of the center of the filament at lag time $t$ and $k=|\vec{k}|$ denotes the wavenumber.
The ISF is directly related to the probability density $\mathbb{P}(\Delta \vec{r},\vec{u},t|\vec{u}_0)$ which measures the probability that an active agent has moved a distance $\Delta\vec{r}$ and changed its orientation from $\vec{u}_0$ to $\vec{u}$ during lag time $t$. The ISF of a single self-propelled agent in free space has been elaborated analytically by solving the associated Fokker-Planck equation in Fourier space
and averaging over swimmer orientations~\cite{Kurzthaler:2016ScRprt} (see SI~\cite{suppl}) 
\begin{equation}
\label{eq:isfanalytic}
F(k,t)\!=\!\frac{1}{2}e^{-D_\perp k^2 t}\sum_{\ell=0}^\infty e^{-A^0_{\ell}t/2\tau_\text{rot}}\left[\int_{-1}^1 \!\diff\eta \  \text{Ps}_\ell^0(\eta)\right]^2.
\end{equation}
Here, $\text{Ps}_\ell^{m}(\eta)\equiv \text{Ps}_\ell^{m}(\eta, R, c)$ are the generalized spheroidal wave functions of order $m$ and degree $\ell$ with eigenvalues $A_\ell^{m} =A_\ell^{m}(R,c)$ which depend on the parameters $R=-2ikv\tau_\text{rot}$ and $c^2=2(D_\parallel -D_\perp) k^2 \tau_\text{rot}$. 

Fig.~\ref{fig:ISF} shows a comparison of the ISFs obtained from simulations and the theoretical predictions of the tube model for different wave numbers $kL$, P{\'e}clet numbers $\text{Pe}$, and number densities  $n^\star$. 
The close agreement over $6$ orders of magnitude in time and $5$ orders of magnitude in space corroborates that the theoretical prediction is in fact an exact result in the highly entangled regime
which  remains valid for all P{\'e}clet numbers, $\text{Pe}=0,\dots, 312$. 

Most prominently, for high P\'{e}clet numbers ($\text{Pe} \gtrsim 31$) the ISFs of highly entangled filaments display oscillations, which is a fingerprint for the persistent swimming motion of an active filament [Figs.~\ref{fig:ISF}(a-b)].  These oscillations can be rationalized by inspecting the general expression of the ISF [Eq.~\eqref{eq:genisf}]. In particular, if the dynamics is dominated by persistent swimming motion, the trajectories can be approximated by $|\Delta \vec{r}(t)|=vt$ leading to  $F(k,t)= \sin(vkt)/vkt$ after averaging over the direction of the needle. Remarkably, we find that this sinc-function (dotted-dashed lines) describes the simulation data nicely for $\text{Pe}\gtrsim 125$ and $kL=4$ [Fig.~\ref{fig:ISF}(b)], yet it fails to capture the dynamics for slower self-propelled agents ($\text{Pe} \lesssim 31$) and their motion at smaller length scales ($kL=50$).


Moreover, the oscillations of the ISF become weaker at smaller length-scales ($kL=50$) [Figs.~\ref{fig:ISF}~(a)] where translational diffusion becomes important. In particular, the tube is effectively leading to an infinite entropic barrier along the perpendicular direction permitting only reptation motion along the long axis. Thus, for a passive filament ($\text{Pe}=0$) we recover an algebraic decay $\sim t^{-1/2}$ of the ISF at intermediate times ($2 \times 10^{-2} \lesssim t/\tau_\text{rot}^0 \lesssim 2 \times 10^{0}$) [Fig.~\ref{fig:ISF}(a)(\emph{inset})], which reflects the sliding motion of the filament from one tube to the next. In fact, this power law is hidden in Eq.~\eqref{eq:isfanalytic} and can be evaluated in a closed-form in the highly entangled regime ($\text{Pe}=0, \tau_\text{rot} \rightarrow \infty$, and $D_\perp \rightarrow 0$), which yields $F(k,t)=\exp(-k^2D_\perp t)/\sqrt{4k^2 (D_\parallel -D_\perp) t/\pi}$ \cite{Leitmann:2016}. 
For intermediate P\'{e}clet numbers ($31 \lesssim \text{Pe} \lesssim 125$), self-propulsion of the neighboring filaments causes a dilation of the effective tube dynamics, thus, speeding up the relaxation process. In particular, the ISF decays to the terminal region exponentially fast [Fig.~\ref{fig:ISF}(a)(\emph{inset})].

At length scales larger than the persistence length of the active agents, i.e. $kv\tau_\text{rot}\lesssim 2\pi$, the swimming direction is randomized and the ISF can be approximated by a relaxing exponential reflecting effective diffusion, $F(k,t)\simeq\exp(-k^2D_\text{eff}t)$~\cite{Kurzthaler:2016ScRprt}.  The ISF for $kL=10^{-3}$ and $\text{Pe}=125$ decorrelates faster for increasing number densities, $n^\star$, which represents a characteristic feature for crowding-enhanced transport [Fig.~\ref{fig:ISF}(c)]. Moreover, the simulation data for $n^\star\lesssim 256$ are described by effective diffusion, while for larger number densities the dynamics are still determined by the directed motion of the active agent as the crowded environment suppresses rotational diffusion.

\paragraph*{Summary and conclusion.--}
Using a minimalistic model system for a solution of active, thin, stiff filaments, we find that transport of its individual components is enhanced by orders of magnitude due to crowding. This finding is indeed counter-intuitive and yet in non-equilibrium systems no fundamental law prevents it. Our study, based on the tube concept pioneered by Doi and Edwards~\cite{Doi:1978tube}, shows that the nature of these transport features relies on the topological constraints of the crowded environment imposed on the swimming direction of the filament. In particular, crowding suppresses the rotational diffusion of self-propelled filaments and, thus, the swimming directions are preserved for long times, leading to an increased effective diffusion.
The tube theory provides a scaling law for the effective diffusivity with respect to crowding and swimming velocity and, thereby, allows for a full spatiotemporal characterization of the active agents. 
Specifically, an exact expression for the ISF unequivocally predicts the dynamics over $6$ orders of magnitude in time and $5$ orders of magnitude in space, valid for all  P\'{e}clet numbers. These findings are remarkable as they represent one of the rare cases where an exact theory can be elaborated for strongly interacting, non-equilibrium systems.

Crowding-enhanced diffusion is a generic mechanism which could occur for any self-driven anisotropic particles in crowded environments. There are various experimental realizations of our model system, both on the macro- and the micro-scale, to verify our predictions. On the macroscale, the dynamics of thin rigid needles, equipped with a self-propelling motor, can be studied under microgravity in high entanglement~\cite{Harth:2018,Harth:2013}. On the micro-scale, motile objects, such as rod-like bacteria, stiff microtubules~\cite{Wu:2017Science,TsangPNAS:2017}, and activated nanotubes, represent potential realizations (see Refs.~\cite{Nagai:2018,Bar:2020} for recent reviews) that could display crowding enhanced diffusion. Moreover, the exact expression for the ISF can be used to extract the motility parameters ($\text{Pe}$, $\tau_\text{rot}$, $D_\parallel$, and $D_\perp$) for strongly interacting non-equilibrium systems via direct comparison with simulations or experimental data obtained, for example, by differential dynamics microscopy~\cite{Cerbino:2008,Martinez:2012}.

Our theory lays the foundation to study the  dynamics of self-driven anisotropic particles in the nematic phase, where the constraint of large-aspect ratio of the single constituents is relaxed. Then, in principle, the use of a tube concept is no longer justified. Computer simulations of passive hard-spherocylinders have been performed along this direction and predict enhanced long-time transport across the isotropic-nematic transition due to orientational 
ordering~\cite{Lowen:1999PRE}, reminiscent of our findings. 
Yet, an analytical theory has not been developed and an extension of our theory for the behavior at the phase transition might explain these simulation data.

Beyond these fundamental interests, our findings propose a potential way to optimize transport of self-driven anisotropic particles in real, densely packed, biological environments or microfluidic devices. The efficient dynamic behavior of these anisotropic particles could be a starting point to design biologically-inspired materials,  e.g., synthetic cells~\cite{Blain:2014,Ganzinger:2019}, or microrobots~\cite{Ceylan:2017mobile}. Moreover, we note that the present work focuses on self-driven stiff filaments, yet, most of the biofilaments present in nature are semiflexible~\cite{Banerjee:2020actin,Eisenstecken:2017,Harnau:1996dynamic,Kroy:1996,Hallatschek:2005,Lang:2018disentangling}. Therefore, a future challenge is to extend our current study to include a finite bending flexibility and identify the disengagement time as a function of swimming speed.

\begin{acknowledgments}
\paragraph*{Acknowledgments.--}
We gratefully acknowledge Sebastian Leitmann and Felix H\"{o}fling for fruitful discussions. This work is supported by the Deutsche Forschungsgemeinschaft (Grant No. LO 418/23-2) and the Austrian Science Fund (FWF) (Grant No. P28687-N27). CK acknowledges support from the FWF via the Erwin Schr{\"o}dinger fellowship (Grant No. J4321-N27).

\end{acknowledgments}


%

\onecolumngrid

\section{Supplemental Material}

\section{Theory for an active Brownian needle in free space} 
\emph{Model.--} We describe the motion of a self-propelled stiff filament in 3D by using the paradigmatic active Brownian particle model~\cite{Kurzthaler:2016ScRprt}. The active needle self-propels at a velocity $v$ along its instantaneous orientation $\vec{u}$, which is subject to rotational Brownian motion with diffusion coefficient, $D_\text{rot}^0$. In addition, the needle performs anisotropic short-time translational diffusion characterized by the diffusivities parallel and perpendicular to its orientation, $D^0_\parallel$ and $D^0_\perp$, respectively.  Thus, the time evolution of the geometric center of the needle $\vec{r}$ and the orientation $\vec{u}$ are described by the Langevin equations,
\begin{align}
\frac{\diff \vec{r}}{\diff t} &=v\vec{u} + \Big[ \sqrt{2 D_\parallel^0} \vec{uu} + \sqrt{2 D_\perp^0} (\mathbb{I} - \vec{uu}) \Big]\boldsymbol\eta, \label{eq:r}\\
\frac{\diff \vec{u}}{\diff t} &= -2D_\text{rot}^0 \vec{u} - \sqrt{2D_\text{rot}^0} \vec{u} \times \boldsymbol\xi. \label{eq:u}
\end{align}
Here, $\boldsymbol{\eta}$ and $\boldsymbol\xi$ are independent, Gaussian white noise processes with zero mean of unit strength, $\langle \eta_i(t)\eta_j(t')\rangle = \langle \xi_i(t)\xi_j(t')\rangle =\delta_{ij}\delta(t-t')$ for $i,j=1,2,3$. Geometrically, Eq.~\eqref{eq:u} can be regarded as Brownian motion of the orientation, $\vec{u}$, on the unit sphere which fulfills the constraint $\diff[\vec{u}^2]/\diff t=0$~\cite{Kurzthaler:2016ScRprt}. We note that $\langle \vec{u}(t)\cdot \vec{u}(0)\rangle = \exp(-2D_\text{rot}^0t)$, which allows introducing the rotational relaxation time as $\tau_\text{rot}^0=1/(2D_\text{rot}
^0)$.  
Equivalently, the motion of the active needle can be described by the probability density $\mathbb{P}    \equiv\mathbb{P}(\Delta\vec{r}, \vec{u},t|\vec{u}_0)$, which measures the probability that the agent has displaced a distance $\Delta\vec{r}$ and changed its orientation from $\vec{u}_0$ to $\vec{u}$ during time $t$. Using standard methods~\cite{Gardiner:2009} the associated Fokker-Planck equation can be derived from the Langevin equations [Eqs.~\eqref{eq:r}-\eqref{eq:u}],
\begin{align}
\partial_t\mathbb{P} = -v\vec{u}\cdot \partial_\vec{r}\mathbb{P}+D_\text{rot}^0\Delta_\vec{u}\mathbb{P}+\partial_\vec{r}\cdot\left[\left(D_\parallel^0\vec{u}\vec{u}+D_\perp^0\left(\mathbb{I}-\vec{u}\vec{u}\right)\right)\cdot\partial_\vec{r}\mathbb{P} \right]   
\end{align}
It is subject to the initial condition, $\mathbb{P}(\Delta\vec{r}, \vec{u},t=0|\vec{u}_0)= \delta(\Delta\vec{r})\delta^{(2)}(\vec{u},\vec{u}_0)$, where the delta function on the surface of the sphere, $\delta
^{(2)}(\cdot,\cdot)$, enforces both orientations to coincide. 

\emph{Analytical solution of the intermediate scattering function.--} We summarize the most important steps from Ref.~\cite{Kurzthaler:2016ScRprt} for deriving the intermediate scattering function (ISF), 
\begin{align}
    F(\vec{k},t)&= \left\langle \exp(-i\vec{k}\cdot\Delta\vec{r})\right\rangle = \int\!\diff^2 u \int\frac{\diff^2 u_0}{4\pi} \ \tilde{\mathbb{P}}(\vec{k}, \vec{u},t|\vec{u}_0).\label{eq:ISF}
\end{align}
Here, we have introduced the Fourier transform of the probability density
\begin{align}
    \tilde{\mathbb{P}}(\vec{k}, \vec{u},t|\vec{u}_0) &= \int \diff^3\Delta r \ \exp(-i\vec{k}\cdot\Delta\vec{r}) \mathbb{P}(\Delta\vec{r}, \vec{u},t|\vec{u}_0),
\end{align}
which obeys the equation of motion
\begin{align}
    \partial_t \tilde{\mathbb{P}}&= -i v \vec{u}\cdot\vec{k} \tilde{\mathbb{P}}+D_\text{rot}^0\Delta_\vec{u} \tilde{\mathbb{P}}-\left[D_\perp^0 k^2+\Delta D \left(\vec{u}\cdot\vec{k}\right)^2\right] \tilde{\mathbb{P}}, \label{eq:characteristic}
\end{align}
with wavenumber $k=|\vec{k}|$ and $\Delta D = D_\parallel^0-D_\perp^0$. Ref.~\cite{Kurzthaler:2016ScRprt} has shown that Eq.~\eqref{eq:characteristic} can be solved by a separation of variables and therefore its solution can be expanded in appropriate eigenfunctions,
\begin{align}
   \tilde{\mathbb{P}}(\vec{k}, \vec{u},t|\vec{u}_0) &=  \frac{1}{2\pi}e^{-k^2D^0_\perp t}\sum_{\ell=0}^{\infty} \sum_{m=-\infty}^\infty e^{i m(\phi-\phi_0)} \text{Ps}_\ell^m(c,R,\eta)\text{Ps}_\ell^m(c,R,\eta_0)e^{-A_\ell^m(c,R) D_\text{rot}^0 t}, \label{eq:p_tilde}
\end{align}
where the orientation has been parametrized by $\vec{u}=\left(\sin\vartheta\cos\varphi, \sin\vartheta\sin\varphi,\cos\vartheta\right)^T$ (and similarly $\vec{u}_0$) and we have abbreviated $\eta = \cos\vartheta$ ($\eta_0 = \cos\vartheta_0$). In particular, $\text{Ps}_\ell^m(c,R,\eta)$ denotes the generalized spheroidal wave function of order $m$ and degree $\ell$, which solves the eigenvalue problem,
\begin{align}
   \left[\frac{\diff}{\diff\eta}\left((1-\eta^2)\frac{\diff}{\diff\eta}\right)+R\eta-c^2\eta^2 -\frac{m^2}{1-\eta^2}+A_\ell^m(c,R)\right]\text{Ps}_\ell^m(c,R,\eta)=0.\label{eq:eigenvalue}
\end{align}
Here, $A_\ell^m(c,R)$ is the associated eigenvalue and the  dimensionless parameters, $c$ and $R$, depend on the motility parameters of the needle via $c^2=\Delta Dk^2/D^0_\text{rot}= 2\Delta Dk^2\tau_\text{rot}^0$ and $R=-i vk/D^0_\text{rot}= -2i vk \tau_\text{rot}^0$.

Inserting Eq.~\eqref{eq:p_tilde} into Eq.~\eqref{eq:ISF} and evaluating the integrals then yields the analytic expression of the ISF
\begin{equation}
\label{eq:isfanalytic}
F(k,t)\!=\!\frac{1}{2}e^{-D_\perp^0 k^2 t}\sum_{\ell=0}^\infty e^{-A^0_{\ell}(c,R)t/(2\tau^0_\text{rot})}\left[\int_{-1}^1 \!\diff\eta \  \text{Ps}_\ell^0(c,R,\eta)\right]^2,
\end{equation}
where only eigenfunctions of order $m=0$ contribute. We note that after integrating over the orientations the ISF depends on the wavenumber $k$ only as the particle motion is isotropic. 

\emph{Numerical evaluation.--} The ISF can be evaluated efficiently numerically by expanding the generalized spheroidal wave functions in terms of Legendre polynomials, $P_j(\eta)$: $\text{Ps}_\ell^m(c,R,\eta)= \sum_{j=0}^\infty d_j^{0\ell}P_j(\eta)\sqrt{(2j+1)/2}$. Thus, the ISF assumes the form, 
\begin{align}
F(k,t) &= e^{-k^2D_\perp^0 t}\sum_{\ell=0}^\infty \left[d_0^{0\ell}\right]^2e^{-A^0_{\ell}(c,R)t/(2\tau^0_\text{rot})}\label{eq:numerics}
\end{align}
where the eigenvalues, $A_\ell^0$,  and coefficients, $d_0^{0\ell}$, are obtained by solving a matrix eigenvalue problem that results from inserting Eq.~\eqref{eq:numerics} into Eq.~\eqref{eq:eigenvalue}. We refer to Ref.~\cite{Kurzthaler:2016ScRprt} for more details. 

\section{Pseudo-Brownian scheme}
We have simulated a collection self-propelled infinitely thin needles via a hard-core interaction potential using event-driven pseudo-Brownian-dynamics simulations~\cite{Scala:2007,Leitmann:2017PRE,Leitmann:2016}, where we employ a collision detection algorithm developed by Frenkel~\cite{Frenkel:1983}. A discrete fixed Brownian time step $\tau_\text{B}$ is used to implement the Langevin equations~[Eqs.~\eqref{eq:r}-\eqref{eq:u}], such that the random pseudo-velocities $\boldsymbol\omega$ and $\vec{v}$ for rotational and translational motion, respectively, are determined at the beginning of every Brownian step. They read
\begin{align}
\begin{split} \label{EqPseudoVelocities}
\boldsymbol\omega &= \sqrt{\frac{2 D_\text{rot}^0}{\tau_\text{B}}}(\mathbb{I} - \vec{uu})\boldsymbol{\mathcal{N}}_\xi ,\\
\vec{v} &= v\vec{u} + \Big[ \sqrt{\frac{2D_\parallel^0}{\tau_\text{B}}} \vec{uu} + \sqrt{\frac{2D_\perp^0}{\tau_\text{B}}} (\mathbb{I} -
\vec{uu}) \Big]\boldsymbol{\mathcal{N}}_\eta, 
\end{split}
\end{align}
where the random variables, $\boldsymbol{\mathcal{N}}_\xi$ and $\boldsymbol{\mathcal{N}}_\eta$, are drawn from a normal distribution with zero mean and unit variance. During a Brownian time step, $\Delta t \in [0,\tau_\text{B}]$, the active needles evolve with constant velocities and collide elastically~\cite{Frenkel:1983} via the following propagation rules:
\begin{align}
\begin{split} \label{EqPropagationRules}
  \vec{u}(t + \Delta t) &= \vec{u}(t)\cos(|\boldsymbol\omega| \Delta t) +
\biggl(\frac{\boldsymbol\omega}{|\boldsymbol\omega|}\times \vec{u}(t)\biggr)\sin(|\boldsymbol\omega|
\Delta t),\\
  \vec{r}(t + \Delta t) &= \vec{r}(t) + \vec{v}\Delta t.
\end{split}
\end{align}



Two active needles may collide when their individual axes are in the same plane~\cite{Frenkel:1983,Huthmann:1999}. This in-plane condition is utilized to predict the next collision time $\tau_c$, see Ref.~\cite{Leitmann:2017PRE,Leitmann:2016} for details.  
At a collision at time $t + \tau_\text{c}$, the translational and angular velocities of the needle,$\vec{v}$ and $\boldsymbol{\omega}$, are determined by enforcing conservation of energy and (angular) momentum. Subsequently, the new translational and rotational velocities, $\vec{v}'$ and $\boldsymbol\omega'$, for the remaining Brownian time interval $[\tau_\text{c},\tau_\text{B}]$ are evaluated as
\begin{align}
\begin{split}
\vec{v}' &= \vec{v} + \Delta p \vec{e}_m,  \\
\boldsymbol\omega' &= \boldsymbol\omega + \frac{m}{I}\Delta p (\vec{r}_\text{coll} \times \vec{e}_m),
\end{split}
\end{align}
where $\Delta p$ is the magnitude of the momentum transfer at the point of contact $\vec{r}_\text{coll} = \vec{r}(t + \tau_\text{c})$ with
respect to the center of the collision partner needle:
\begin{align} 
\Delta p = -2\frac{\vec{v}\cdot\vec{e}_m +
\boldsymbol\omega\cdot(\vec{r}_\text{coll}\times\vec{e}_m)}{1+\frac{m}{I}(\vec{r}_\text{coll}\times\vec{e}_m)^2} .
\end{align}
In fact, the momentum transfer for smooth needles occurs perpendicular to the orientation of both collision partners and is directed along $\vec{e}_m =\vec{u}(t+\tau_c) \times \vec{u}_c/|\vec{u}(t+\tau_c) \times \vec{u}_c|$, where $\vec{u}_c$ denotes the orientation of the collision needle. Furthermore, the ratio of mass $m$ and inertia $I$ can be related to the short-time diffusion coefficients~\cite{Leitmann:2016} by 
\begin{align}
\frac{m}{I} = \frac{D_\text{rot}^0}{D_\perp^0},
\end{align}
which prohibits an average flow of energy between the rotational and translational degrees of freedom of the collision partner needle.

\section{Mean-square displacement for different P\'eclet numbers}
\begin{figure}[htp]
	\includegraphics[width = \linewidth]{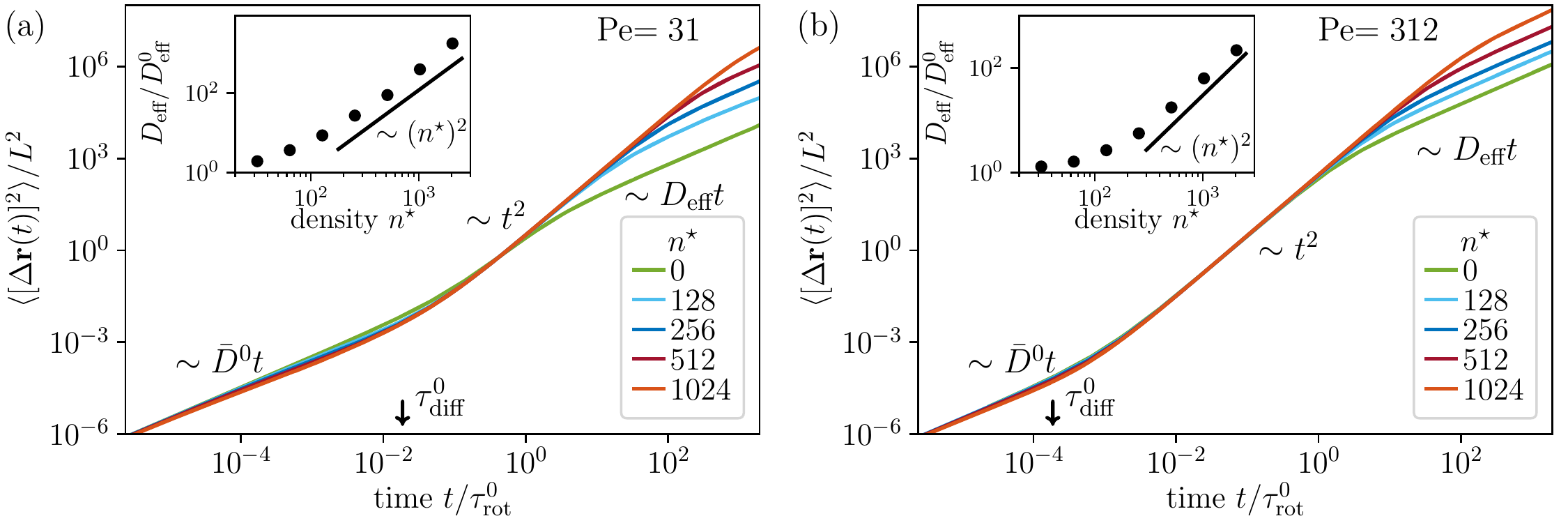}
		\caption{\label{fig:enhanced_diffusion_Pe312}
Mean-square displacement $\langle[\Delta\mathbf{r}(t)]^2\rangle$ of a self-propelled stiff filament suspended in needle solutions with different number densities $n^\star=nL^3$. Panels (a) and (b) show data for different P\'eclet numbers, Pe$=31$ and Pe$=312$, respectively.
(\emph{Insets}) Effective diffusion coefficient $D_\text{eff}$ rescaled by the effective diffusivity in a free environment, $D_\text{eff}^0$, as a function of the number density $n^\star$. The solid black line indicates a quadratic increase.
	}
\end{figure}

The mean-square displacement (MSD) for different P{\'e}clet numbers, Pe$=31,312$, is shown in Fig.~\ref{fig:enhanced_diffusion_Pe312} for different number densities~$n^\star$. The MSDs display a diffusive behavior at times $t\lesssim \tau_\text{diff}:=\bar{D}^0/v^2$, which is followed by a regime of directed motion, $\sim t^2$. In fact, this directed regime is extended by roughly two orders of magnitude in time for Pe=$312$ where the cross-over time is $\tau_\text{diff}\simeq 2\times 10
^{-4}\tau_\text{rot}^0$ [Fig.~\ref{fig:enhanced_diffusion_Pe312}(b)] compared to smaller P{\'e}clet number, Pe$=31$, with $\tau_\text{diff}\simeq 2\times 10
^{-2} \tau_\text{rot}^0$ [Fig.~\ref{fig:enhanced_diffusion_Pe312}(a)]. For comparison, the cross-over time for Pe$=125$, as shown in Fig.~1 in the main text, is $\tau_\text{diff}\sim 10^{-3}\tau_\text{rot}^0$. At long times, $t\gtrsim\tau_\text{rot}^0$ the MSDs become diffusive again and show enhanced effective diffusion by increasing the number density $n^\star$.

\section{Orientational correlation function}
\begin{figure}[htp]
	\includegraphics[width = \linewidth]{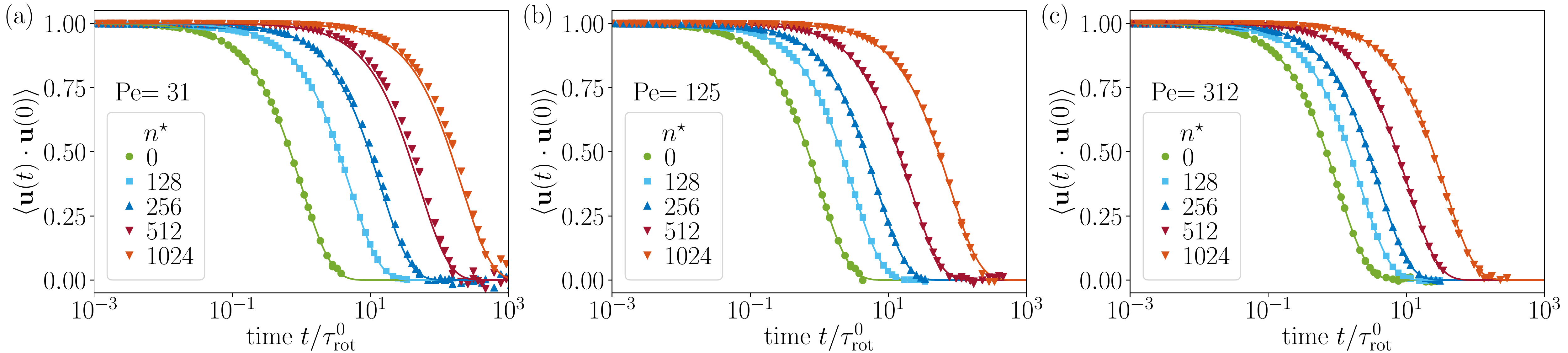}
		\caption{\label{fig:ocf}
Orientational correlation function, $\langle\vec{u}(t)\cdot\vec{u}(0)\rangle$, of a self-propelled stiff filament suspended in needle solutions with different number densities $n^\star=nL^3$. Panels (a-c) show data for different P\'eclet numbers, Pe$=31, 125,$ and $312$, respectively.
Symbols correspond to simulation results and lines to the fit of the orientational correlation function to a relaxing exponential, $\exp(-t/\tau_\text{rot})$, where $\tau_\text{rot}$ denotes the rotational relaxation time.}
\end{figure}

Figure~\ref{fig:ocf} depicts that the orientational correlation functions extracted from simulations display an exponential decay and decay slower for increasing number densities, $n^\star$. In particular, we find that the orientational correlation functions can be well described by a relaxing exponential, $\langle\vec{u}(t)\cdot \vec{u}(0)\rangle=\exp(-t/\tau_\text{rot})$, where we have introduced the rotational relaxation time $\tau_\text{rot}$. For a free environment, $n
^\star=0$, the rotational relaxation time reduces to the short-time rotational relaxation time, $\tau_\text{rot}=\tau_\text{rot}^0$, as noted earlier. Comparison to the simulation results for higher number densities, $n^\star$, then allows quantitative extraction of the rotational relaxation times, which are shown in Fig.~2(c) of the main text. 

\section{Tube diameter}
\begin{figure}[htp]
	\includegraphics[width=0.9\linewidth]{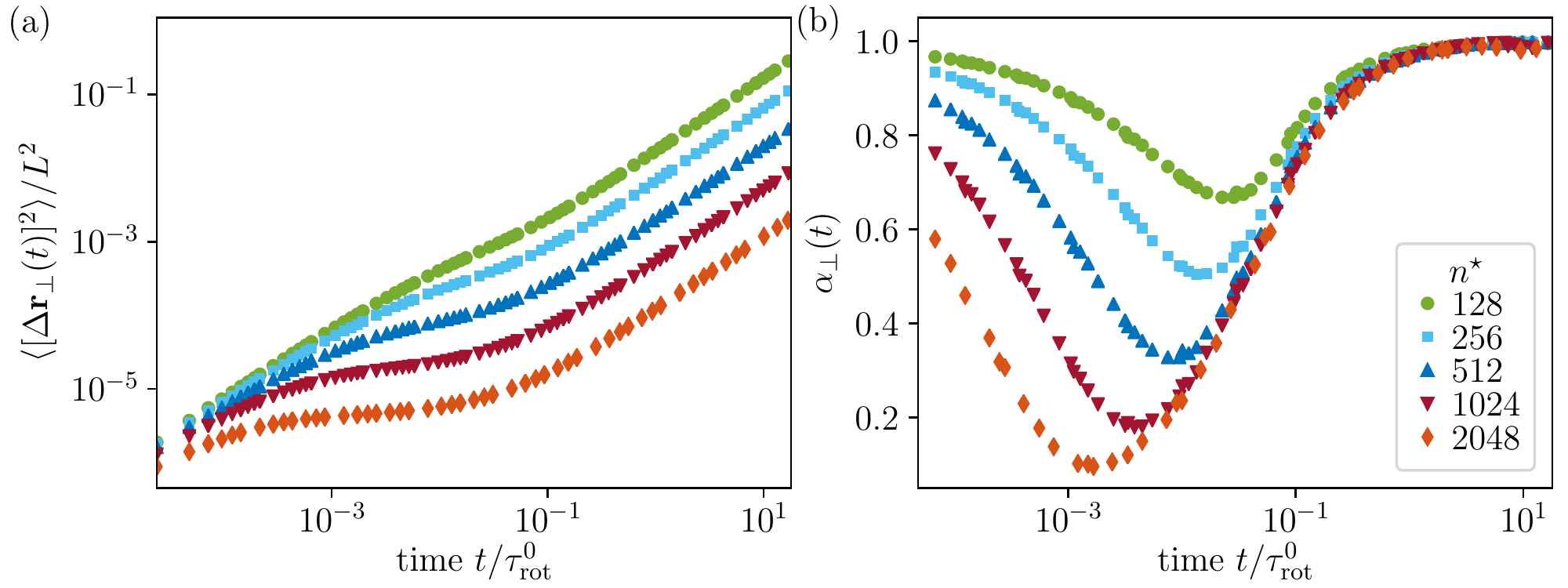}
		\caption{\label{fig:tube_scaling}
(a) The mean-square displacement perpendicular to the needle axis ($\langle[\Delta\vec{r}_\perp(t)]^2\rangle$) in the comoving frame as a function of density $n^\star$ for $\text{Pe}=125$. (b) The corresponding local exponent $\alpha_\perp(t)=\diff \ln(\langle[\Delta\vec{r}_\perp(t)]^2\rangle)/ \diff \ln(t)$ as a function of density $n^\star$ for $\text{Pe}=125$. 
	}
\end{figure}
The tube is effective once the perpendicular fluctuations are limited. This is reflected by an extended plateau of the mean-square displacement along the direction perpendicular to the needle, $\langle\left[\Delta \vec{r}_\perp(t)\right]^2\rangle$, at intermediate times, see Fig.~\ref{fig:tube_scaling}(a). To realize the tube diameter scaling, we consider a cylindrical tube of length $L$ and diameter $R$. The average number of stiff filaments intersecting the cylinder can be estimated as $nLS_c(R)$, where $n$ is the number of needles per unit volume and $S_c(R)=2\pi L R$ is the surface area of the cylinder. Once $R \sim d$, the average number of filaments within the tube should be of order $\mathcal{O}(1)$, i.e., $2\pi d nL^2 \sim 1$, which implies $d \sim 1/nL^2$~\cite{DoiEdwards:1988}.

To quantify the tube diameter $d$ from simulations, we measure the local exponent,
\begin{align}
\alpha_\perp(t)=\frac{\diff \ln(\langle[\Delta\mathbf{r}_\perp(t))^2\rangle]}{\diff \ln(t)},
\end{align}
as a function of time and for different densities $n^\star$, see Fig.~\ref{fig:tube_scaling}(b). We find that the local exponent is getting more and more suppressed at intermediate times for increasing density $n^\star$. Subsequently, we estimate the time $\tau_\perp$ corresponding to the minimum in $\alpha_\perp(t)$ and define the tube diameter  via
\begin{equation}
\langle[\Delta\mathbf{r}_\perp(\tau_\perp)]^2\rangle=d^2.
\end{equation}
We find that the predicted scaling, $d/L \sim 1/(nL^3) = (n^\star)^{-1}$, remains valid for all P\'eclet numbers, see Fig.~2(b) in the main text. However, it shifts to larger number densities with increasing P\'eclet number.

\end{document}